\documentclass[epj]{svjour}

\newcommand{\Vista}{{\sc Vista}}
\newcommand{\Sleuth}{{\sc Sleuth}}
\usepackage{graphicx}
\usepackage{fancyhdr}
\usepackage{amsmath}
\usepackage{amssymb}
\def\makeheadbox{{
 }}
\newcommand {\abs}[1]{\left| #1 \right|}
\def \pmiss {{\ensuremath{{\,/\!\!\!p}}}}

\def\mytitle{My title} 
\def\myauthors{My name}  
\def\mytype{My type of session}
\def\mysession{My session}

\def\mytitle{Vista at CDF} 
\def\myauthors{Georgios Choudalakis}    
\def\mytype{Contributed Talk}    
\def\mysession{Alternatives}

\begin{document}
\title{Vista at CDF}
\subtitle{Results of a model-independent search for new physics in 927~pb$^{-1}$ at CDF}\author{Georgios Choudalakis\thanks{http://www.mit.edu/$\sim$gchouda/, gchouda@mit.edu}%
}                     
%
%
\institute{Massachusetts Institute of Technology, for the CDF collaboration}
%
\date{}
\abstract{
A global, model-independent search for high-$p_T$ exotic phenomena is presented using 927~pb$^{-1}$ of CDF II data. The search algorithms employed in this analysis are \Vista\ and \Sleuth.  These proceedings focus on \Vista, including a description of the method and a summary of results.
} 
\maketitle
\section{Introduction}
\label{intro}
A model-independent search for new physics is presently well motivated.  While the $W$ boson, $Z$ boson, and top quark represented very specific predictions of an already well established Standard Model, and were therefore nearly guaranteed targets, the exciting physics expected to lie beyond the Standard Model may assume any of a number of different forms.  Even within the {{\em{Minimal}} Supersymmetric Standard Model, different points in the 105 dimensional parameter space  correspond to a large array of possible signatures.  

Performing a global search requires a concrete and practical strategy.  The approach taken by most searches of more limited scope involves selecting a proposed model of new physics from the existing literature and searching for its signature in data collected at the energy frontier.  An alternative approach takes a somewhat broader view, imposing less restrictive assumptions as to what the first signature of new physics may be.  Rather than defining traditional ``control'' and ``signal'' regions, all regions of the data are considered to potentially harbor the first sign of new physics.  Simultaneously, all regions contribute information used collectively to constrain the predicted Standard Model background.  The Standard Model prediction consists entirely of Monte Carlo events (except for estimation of non-collision backgrounds, including cosmic rays and beam halo, modeled using data with few reconstructed tracks).  This prediction is compared with CDF data in all high-$p_T$ final states, in a large number of kinematic variables constructed from 4-vector quantities.  A minimal set of well motivated correction factors enters the calculation of the Standard Model background, with values adjusted under external constraints to minimize global (large-scale) disagreement with data.  Discrepancies that persist in spite of efforts to achieve agreement by reasonable adjustments of the correction model may motivate a discovery claim.  

\section{The method}
\label{sec:method}

The CDF detector, described elsewhere \cite{cdfDescription}, records collisions of protons and antiprotons at $\sqrt{s}=1.96$~TeV.  The Monte Carlo events composing the Standard Model prediction are generated primarily using {\sc{pythia}}~\cite{Pythia}, {\sc{herwig}}~\cite{Herwig} and {\sc{madevent}}~\cite{MadEvent}.  After generation, the Standard Model events pass through a {\sc{geant}}-based simulation of the CDF detector.

In each event, energetic and isolated ``objects'' --- electrons ($e$), muons ($\mu$), photons ($\gamma$), taus ($\tau$), non-$b$-quark jets ($j$), $b$-tagged jets ($b$), and missing transverse momentum ($p\!\!\!/_T$) --- are identified with sufficiently large transverse momentum ($p_T > 17$~GeV).  Roughly two million data events with one or more sufficiently high-$p_T$ objects are included in the analysis.  

Events so selected are partitioned into exclusive final states according to reconstructed final state objects.  The $e^+e^-$ final state thus consists of all events containing exactly one positron, one electron, and no other reconstructed object.  The partitioning is orthogonal, with each event associated with one and only one final state.  Possible final states are defined algorithmically, and are dynamically created to accommodate all events:  an observed event with seventeen muons would prompt the creation of a corresponding final state, which would then be included in subsequent analysis.  Exclusive final states allow an algorithmic specification of a finite set of kinematic variables that make sense for all events in each final state.  Applicable variables in each final state include object transverse momenta, polar and azimuthal angles, angles between pairs of objects, masses of all object combinations, and a number of additional specialized variables.  With data and Standard Model background events partitioned by the same rule, the search for discrepancies can proceed separately in each final state.

Standard Model Monte Carlo events are adjusted by a minimal set of theoretical and experimental correction factors that include the integrated luminosity, one $k$-factor\footnote{Here a $k$-factor is defined as the ratio of the actual (unknown) Standard Model cross section and the (known) leading order cross section.} for each of roughly twenty processes, approximately twenty object (mis)identi\-fi\-ca\-tion probabilities, and four online trigger efficiencies.  With only these 44 correction factors, a comparison is made between data and Standard Model prediction in over three hundred exclusive final states and over ten thousand kinematic distributions.  The goal of the \Vista\ correction model is not necessarily a perfectly accurate estimation of the Standard Model background in all final states, but rather an estimation reliable enough to indicate if a signal of new physics is present.

A global fit determines the values of the correction factors, using simultaneously information from all data entering the analysis, together with external information where applicable.  The fit minimizes a binned $\chi^2$, which is a function of the correction factors ($\vec{s}$):
\begin{equation}
\chi^2(\vec{s})= \sum_{k \in {\rm bins}} \frac{\left({\rm Data}[k] - {\rm SM}[k]\right)^2}{\delta{\rm SM}[k]^2 + \sqrt{{\rm SM}[k]}^2} + \chi^2_{\rm constraints}.
\end{equation}
Bins $k$ loosely correspond to exclusive final states, with additional division in object transverse momentum ($p_T$) and pseudorapidity ($\eta$).  In each bin $k$, the Standard Model background (${\rm SM}[k]$) is a function of the values of the correction factors $\vec{s}$, depending on the overall integrated luminosity, $k$-factors of contributing Standard Model processes, object identification efficiencies and misidentification rates, and trigger efficiencies.  The term $\chi^2_{\rm constraints}$ increases if a correction factor assumes a value different from the value preferred by external sources of information, such as an NLO calculation.  

With correction factor values globally determined, a comparison is performed between data and Standard Model prediction to highlight any remaining significant discrepancies.  Discrepancies may prompt additional refinement of the Standard Model prediction or detector response if such adjustment is not inconsistent with existing experimental knowledge.  The global fit and search for remaining discrepancies is repeated after each adjustment, testing the consistency of the adjustment with all available high-$p_T$ data.  Iteration occurs until either a clear case for new physics can be made, or there remain no discrepancies that may motivate such a case.  Judgement is used to implement only physically motivated improvements in this procedure, rather than {\em ad hoc} modifications that remove discrepancies without physical reasoning.  In this spirit, emphasis on physical understanding within \Vista\ has resulted in a quantitative and unified understanding of the underlying physics responsible for the misidentification of jets as electrons, muons, taus, and photons at CDF.

\section{Results of the \Vista\ comparison}
\label{sec:results}
The first 927 pb$^{-1}$ of CDF II data populate 344 exclusive final states.  The first \Vista\ statistic quantifies the difference between the observed and predicted populations of these final states (Fig.~\ref{fig:summary}-a). For each final state, the Poisson probability that the expected population would fluctuate up to or above (or down to or below) the observed number of events is calculated.  A trials factor associated with examining 344 final states reduces the significance of the observed discrepancies.  The largest population discrepancy, corresponding to a 2.3$\sigma$ deficit of data after this trials factor is accounted for, is not statistically significant.

\begin{figure}
$\begin{array}{c}
\includegraphics[width=0.45\textwidth,angle=0]{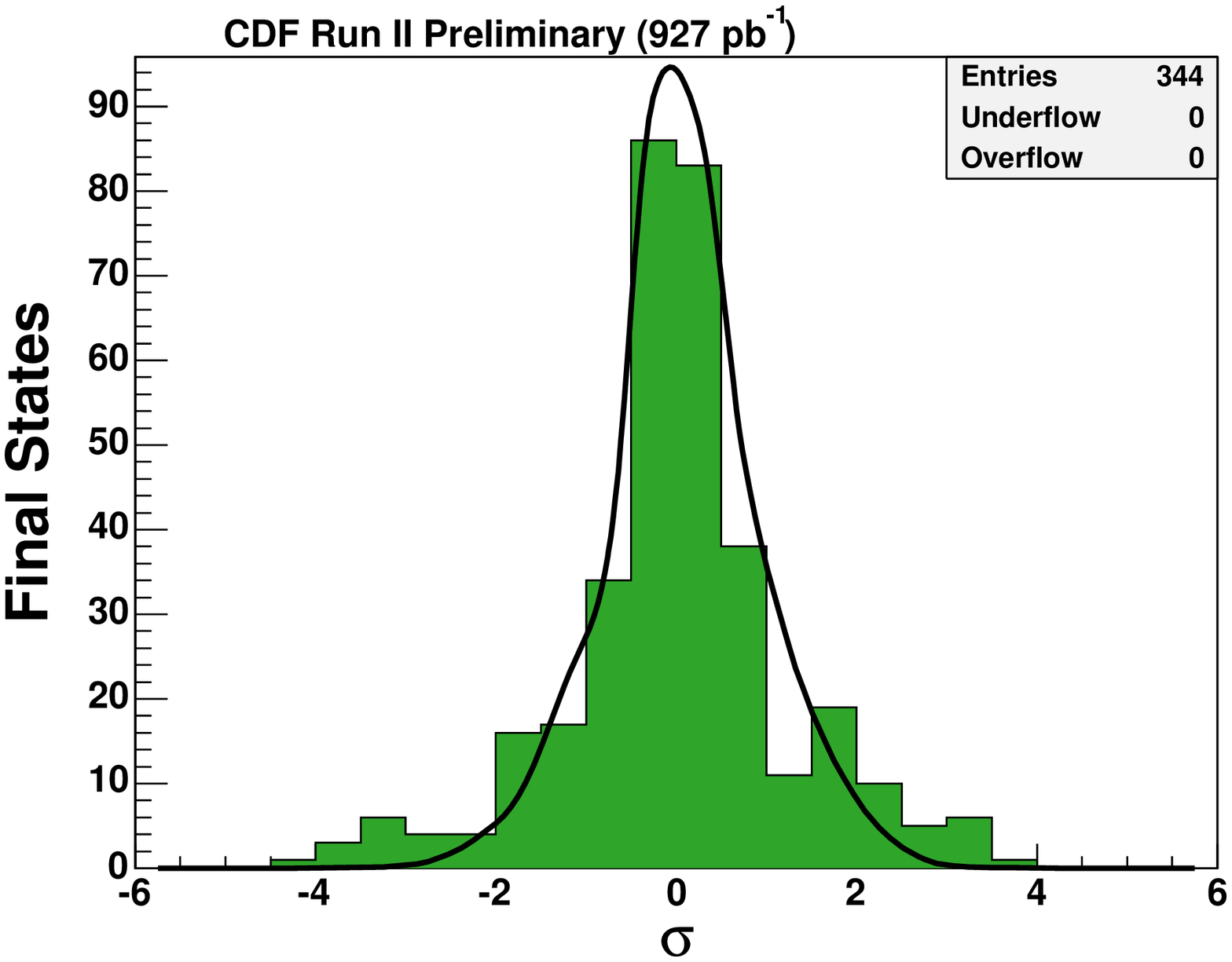} \\
(a) \\
\includegraphics[width=0.45\textwidth,angle=0]{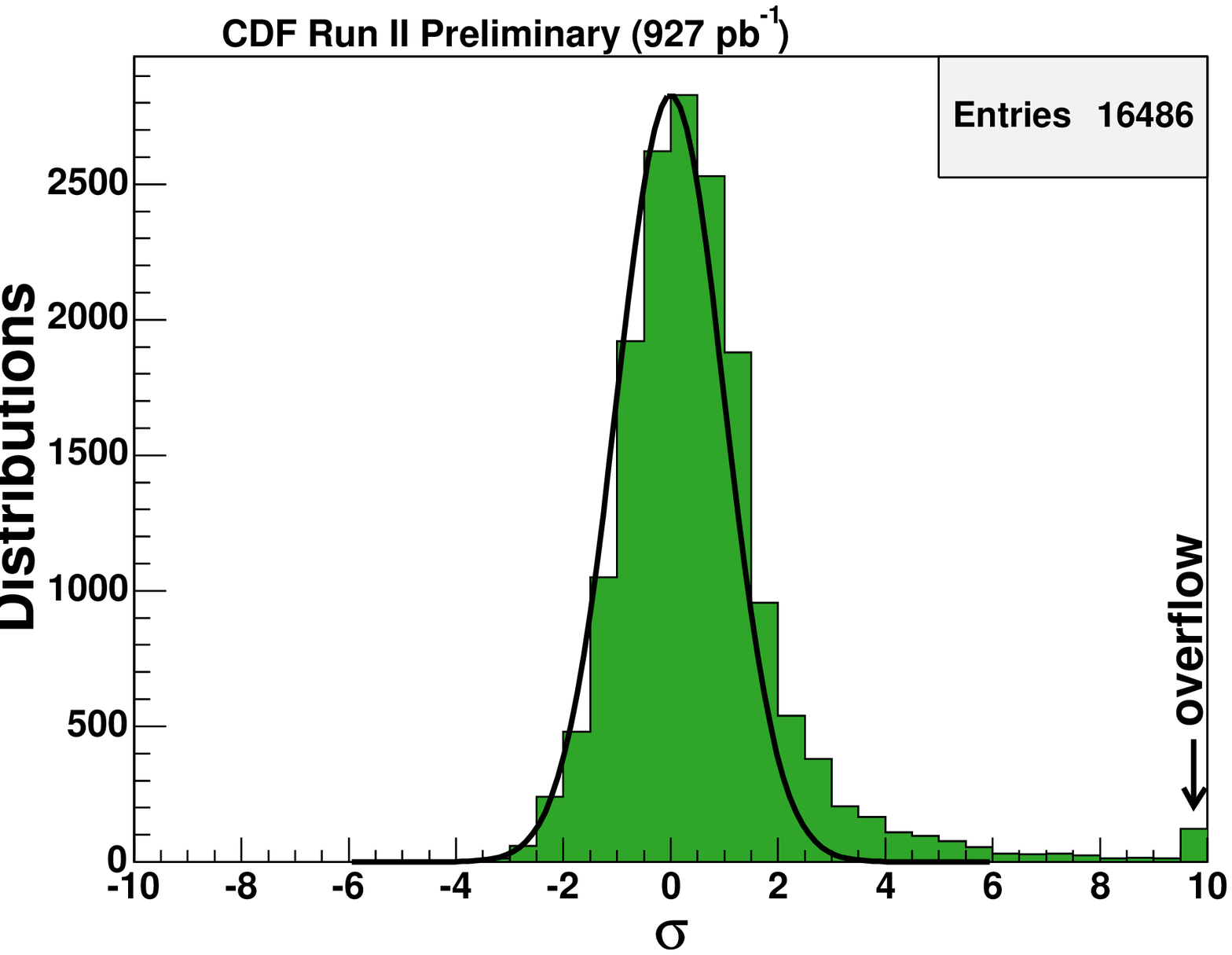} \\
(b)
\end{array}$
\caption{{\bf (a)} \Vista\ population discrepancies, quantifying the difference between the number of events observed and predicted in each of the 344 \Vista\ final states considered.  Final states containing more data than Standard Model prediction populate the right side of the distribution, while final states containing fewer data events populate the left.  {\bf (b)} \Vista\ shape discrepancies, quantifying the difference in shape between data and Standard Model prediction in 16,486 kinematic variables.  The horizontal axis ranges from agreement to disagreement in shape from left to right.  In both (a) and (b) the black curve is the expected distribution, obtained by drawing pseudo data from the Standard Model background.  The horizontal axis in both (a) and (b) represents statistical significance, in units of standard deviations ($\sigma$), before accounting for the associated trials factor (see text). }
\label{fig:summary}
\end{figure}

In addition to total final state populations, \Vista\ examines shapes of kinematic distributions.  In each final state, \Vista\ algorithmically produces the distributions of a large number of potentially informative kinematic variables.  The total number of distributions considered in all final states is 16,486.  The Kolmogorov-Smirnov (KS) test is used to evaluate the agreement of data with the Standard Model prediction in each kinematic variable.  The distribution of these KS probabilities, converted into units of standard deviations ($\sigma$), is shown in Fig.~\ref{fig:summary}-b.  A trials factor equal to the number of distributions considered reduces the significance of any individual observed shape discrepancy. 

While the number of events observed in each final state does not result in a statistically significant discrepancy that might motivate a new physics claim, consideration of the shapes of kinematic distributions result in a few hundred shape discrepancies that remain statistically significant even after accounting for the associated trials factor.  These shape discrepancies can be generally categorized as manifestations of the modeling of the intrinsic transverse boost of the event, and modeling the angular separation between subleading jets.    

Accurate modeling of the intrinsic boost ($k_T$ kick) of events produced at a hadron collider is a long-standing problem.  The symptomatic \Vista\ distributions include the total energy visible in the detector but not clustered into any specific reconstructed object; missing transverse energy in events where this missing transverse energy is not significant, such as in dijet, diphoton, and $Z$ production; the projection of the vector summed momenta of all reconstructed objects along and perpendicular to the thrust axis in the event; and other related variables.  Although a satisfactory solution to this problem has not yet been obtained, \Vista\ currently supplies a reasonably comprehensive catalog of relevant experimental information from $p\bar{p}$ collisions.

The mismodeling of the angular separation between subleading jets is shown most clearly in Fig.~\ref{fig:3j}.  Many other discrepant distributions derive from the effect shown in this low-$p_T$ final state, consisting of one central ($\abs{\eta}<1$) jet having $p_T>40$~GeV and two additional reconstructed jets with $\abs{\eta}<2.5$ and $p_T>17$~GeV.  Derived discrepancies include the masses of individual jets, where jets in data are observed to be systematically more massive than jets from the {\sc Pythia} Standard Model prediction.  Although no complete, quantitative understanding has yet been achieved, pursuit of a showering-based explanation is ongoing.

\begin{figure}
\centering
\includegraphics[height=0.45\textwidth,angle=270]{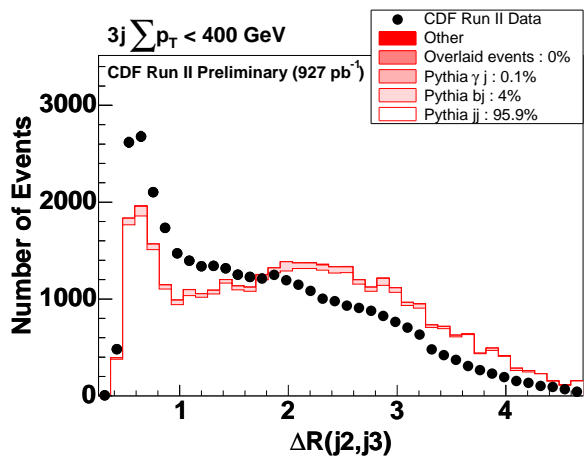} 
\caption{A shape discrepancy highlighted by \Vista\ in the final state consisting of exactly three reconstructed jets with $\abs{\eta}<2.5$ and $p_T>17$~GeV, and with one of the jets satisfying $\abs{\eta}<1$ and $p_T>40$~GeV.  The discrepancy is clearly statistically significant, with statistical error bars smaller than the size of the data points.  The vertical axis shows the number of events per bin, with the horizontal axis showing the angular separation ($\Delta R=\sqrt{\Delta\eta^2+\delta\phi^2}$) between the second and third jets, where the jets are ordered according to decreasing transverse momentum.  The region $\Delta R(j_2,j_3)\gtrsim2$ is populated primarily by initial state radiation, and here the Standard Model prediction can to some extent be adjusted; the region $\Delta R(j_2,j_3)\lesssim2$ is dominated by final state radiation, the description of which is constrained by data from LEP\,1.}
\label{fig:3j}
\end{figure}

\section{Conclusion}
\label{sec:conclusion}

These proceedings have motivated and briefly outlined the \Vista\ global analysis, together with the result obtained on 927~pb$^{-1}$ of CDF II data.  This analysis represents the first model-independent search in hadron-hadron collider data of this scope, including 16,486 kinematic variables in 344 populated exclusive final states defined by seven reconstructed objects ($e$, $\mu$, $\tau$, $\gamma$, $j$, $b$, $\pmiss$).  This result should not be construed as having proven that there is no new physics hiding in the Tevatron data; merely that this analysis has not revealed an indication of a discrepancy appearing to motivate a new physics claim.  New physics above the electroweak scale appearing with low cross section represents a more specific target of \Sleuth~\cite{sleuthPRL,sleuthPRD1,sleuthPRD2}, discussed in a companion proceedings~\cite{sleuth}.


\begin{thebibliography}{999}
%
%
\bibitem{cdfDescription}
CDF Collaboration, Phys.~Rev.~{\bf D 71} (2005) 032001.
\bibitem{Pythia}
T.~Sjostrand \emph{et al.}, Comput.~Phys.~Commun. {\bf 135} (2001) 238, PYTHIA 6.203.
\bibitem{Herwig}
G. Corcella \emph{et al.}, J.~High Energy Phys.~0101 (2001) 010.
\bibitem{MadEvent}
T. Stelzer and W.F. Long, Comput.~Phys.~Commun.~{\bf 81}, 357-371 (1994).
\bibitem{sleuthPRL}
D0 collaboration, Phys.~Rev.~Lett. 86 (2001) 3712--3717.
\bibitem{sleuthPRD1}
D0 collaboration, Phys.~Rev.~D62 (2000) 092004.
\bibitem{sleuthPRD2}
D0 collaboration, Phys.~Rev.~D64 (2001) 012004.
\bibitem{sleuth}
G.~Choudalakis, companion (SUSY 07) proceedings.
\end{thebibliography}
\end{document}